\documentclass[aps,superscriptaddress,showpacs,twocolumn]{revtex4}
\usepackage{graphicx}
\usepackage{color}
\usepackage{amssymb}
\usepackage{amsmath}

\newcommand{\bk}{{\bf k}}

\newcommand{\kB}{k_{\mathrm{B}}}

\begin{document}

\title{Comment on ``Variation of the superconducting transition
   temperature of hole-doped copper oxides''.}

\author{G. G. N. Angilella}
\email{Giuseppe.Angilella@ct.infn.it}
\affiliation{Dipartimento di Fisica e Astronomia, Universit\`a di
   Catania,\\ and Istituto Nazionale per la Fisica della Materia,
   UdR di Catania,\\ Via S. Sofia, 64, I-95123 Catania, Italy}
\author{R. Pucci}
\affiliation{Dipartimento di Fisica e Astronomia, Universit\`a di
   Catania,\\ and Istituto Nazionale per la Fisica della Materia,
   UdR di Catania,\\ Via S. Sofia, 64, I-95123 Catania, Italy}
\author{A. Sudb\o}
\affiliation{ Department of Physics, Norwegian University of Science
   and Technology, N-7491 Trondheim, Norway}

\date{\today}

\begin{abstract}
\medskip
We point out the incorrect derivation of the gap equation in X.-J. Chen
   and H. Q. Lin [Phys. Rev. B {\bf 69}, 104518 (2994)] within the
   interlayer tunneling (ILT) model for multilayered cuprates. 
There, the \emph{local} structure in $\bk$-space of the ILT effective
   interaction has not been taken into due account when the ILT model
   is generalized to the case of $n$ layers per unit cell. 
This is a specific characteristic of the ILT model that, apart from giving
   rise to a highly nontrivial $\bk$-dependence of the gap function, is
   known to enhance the critical temperature $T_c$ in a natural way.
As a consequence, we argue that Chen and Lin's results cannot be
   employed, in their present form, for a quantitative interpretation
   of the high-pressure dependence of $T_c$ in Bi-2212, as is done by
   X.-J. Chen \emph{et al.} [{\tt cond-mat/0408587}, to appear in
   Phys. Rev. B].
Moreover, when the generalization of  X.-J. Chen \emph{et al.} [{\tt
   cond-mat/0408587}] is applied to the case $n=2$, it fails to
   reproduce the original ILT gap equation. 
However, a more careful analysis of the ILT model for multilayered
   cuprates, taking into account the nonuniform hole distribution
   among inequivalent layers, has been earlier suggested to describe 
   the observed pressure dependence of $T_c$ in homologous series of
   high-$T_c$ cuprates.
\\
\end{abstract} 

\maketitle


In Ref.~\onlinecite{Chen:04}, Chen and Lin reconsider the dependence
   of $T_c$ on doping and on the number of layers in a homologous series of
   multilayered high-$T_c$ cuprates within the interlayer tunneling
   (ILT) model \cite{Chakravarty:93}. 
However, in deriving their gap equation, Chen and Lin erroneusly
   neglect the intrinsic \emph{local} structure in momentum ($\bk$)
   space of the effective ILT coupling.
This is a specific characteristic of the ILT model, which is known to give
   rise to highly nontrivial features in the $\bk$-dependence of the
   gap function already for a bilayer complex
   \cite{Angilella:99,Angilella:00}.
Moreover, a local term in the gap equation has been shown to provide a
   lower bound for $T_c$ at all dopings, which is the precise way in
   which the ILT mechanism enhances $T_c$ \cite{Angilella:99}.
The consequences of such an incorrect analysis of the ILT model are
   both qualitative and quantitative.
Therefore, the recent use of Chen and Lin's results to interpret the
   high-pressure dependence of $T_c$ in Bi-2212 \cite{Chen:04a} can be
   questioned.
In this context, we point out that a more careful analysis of the ILT
   model for layered cuprates has been presented elsewhere
   \cite{Sudboe:94c}, and successfully applied to study the pressure
   dependence of $T_c$ in homologous series of layered cuprates, by
   explicitly taking 
   into account the inhomogeneous hole-doping in inequivalent
   layers \cite{Angilella:99b,Wijngaarden:99}.

Superconductivity in the high-$T_c$ layered cuprates is characterized
   by \emph{(i)} a non-monotonic dependence of $T_c$ on the overall hole-doping
   $\delta$; \emph{(ii)} a monotonic increase of $T_c$ with the number of
   layers $n$, for moderately low $n$ ($n\lesssim 3$).
While \emph{(i)} is a generic consequence of the quasi-bidimensional nature
   of these compounds (see \emph{e.g.} Ref.~\onlinecite{Angilella:01}), the
   latter fact has suggested that coherent tunneling of
   superconducting pairs between adjacent CuO$_2$ layers may
   considerably enhance $T_c$ \cite{Chakravarty:93}.
Within the ILT model, it is postulated that strong in-plane
   correlations forbid coherent hopping of single
   particles between adjacent CuO$_2$ planes.
Such a restriction is removed when accessing the superconducting
   state, where interlayer Josephson tunneling of Cooper pairs is
   allowed.
This results in a net gain in kinetic energy, as compared to the
   normal state.
Thus, within the ILT model, superconductivity is stabilized \emph{via} a
   kinetic mechanism, as opposed to conventional BCS
   superconductivity, where the enhancement in kinetic energy is
   overcompensated by a reduction in the potential energy
   \cite{Chakravarty:98a}.

However, after its original formulation more than a decade ago
   \cite{Chakravarty:93}, the relevance of the ILT mechanism at least
   for single-layer cuprates has been called into question by
   experiments \cite{Anderson:98,Moler:98}.
Recently, Chakravarty \emph{et al.} \cite{Chakravarty:04} have revived
   the ILT model in connection with multilayered cuprates.
There, ILT needs not be the sole source of superconducting
   condensation energy.
Charge carriers require a `seed' in-plane interaction to form Cooper
   pairs in a given symmetry channel, before they can actually tunnel
   between adjacent layers \cite{Angilella:99}.
Such in-plane interaction would then provide the missing condensation
   energy \cite{Chakravarty:04}.

Moreover, it has been suggested that the competition with a `hidden'
   order parameter, such as a $d$-density-wave (dDW)
   \cite{Chakravarty:01}, could be responsible for the downturn of $T_c$
   with $n$, for $n\gtrsim 3$.
Indeed, in multilayered cuprates, due to the different proximity to
   the `charge reservoir' blocks, experiments \cite{Trokiner:91} as
   well as density functional theory calculations
   \cite{Ambrosch-Draxl:04} revealed a nonuniform hole-content
   distribution between inner and outer layers.
Since this usually places inner (outer) layers in the underdoped
   (overdoped) region of the cuprate phase diagram, competition with
   the dDW order would be stronger in inner layers than in outer
   layers, thus depressing $T_c$ with increasing $n$.
Hydrostatic pressure could then be used to tune both the overall
   hole-content content and its distribution among inequivalent
   layers, thus inducing an `exchange of roles' between inner and outer
   layers with respect to the onset of superconductivity
   \cite{Angilella:99b}, which is observed as `kinks' in the
   pressure dependence of $T_c$ in layered cuprates
   \cite{Wijngaarden:99}.

The effective Hamiltonian considered by Chen and Lin \cite{Chen:04}
   (see also \cite{Angilella:99,Angilella:99b}) is
\begin{eqnarray}
H &=& \sum_{\ell \bk\sigma} \xi_\bk c_{\bk\sigma}^{\ell\dag}
   c_{\bk\sigma}^\ell - \sum_{\ell\bk\bk^\prime} V_{\bk\bk^\prime}
   c_{\bk\uparrow}^{\ell\dag} c_{-\bk\downarrow}^{\ell\dag}
   c_{-\bk^\prime \downarrow}^\ell c_{\bk^\prime \uparrow}^\ell
\nonumber\\
&&+ \sum_{\langle\ell\ell^\prime \rangle} \sum_\bk T_J (\bk)
   c_{\bk\uparrow}^{\ell\dag} c_{-\bk\downarrow}^{\ell\dag}
   c_{-\bk\downarrow}^{\ell^\prime} c_{\bk\uparrow}^{\ell^\prime} ,
\label{eq:H}
\end{eqnarray}
where $\xi_\bk$ is the in-plane quasiparticle dispersion measured with
   respect to the chemical potential $\mu^\ell$ ($\mu^\ell \equiv \mu$
   for all layers, in Ref.~\onlinecite{Chen:04}), and
   $c_{\bk\sigma}^{\ell\dag}$ 
   is a quasiparticle creation operator with wave-vector $\bk$ and spin
   $\sigma$ on layer $\ell$.
It should be emphasized that in Eq.~(\ref{eq:H}) the first interaction
   term ($V_{\bk\bk^\prime}$) pertains to a single layer and governs the
   overall symmetry of the order parameter (\emph{i.e.}, $d$-wave, if
   $V_{\bk\bk^\prime} = V g_\bk g_{\bk^\prime}$, with $g_\bk =
   \frac{1}{2} (\cos   k_x - \cos k_y )$) \cite{Sudbo:95b}, while the
   second term applies to adjacent layers 
   $\langle\ell\ell^\prime \rangle$, and is \emph{local} in
   $\bk$-space, with $T_J (\bk) = \frac{1}{16} T_J (\cos k_x - \cos
   k_y )^4$ \cite{Chakravarty:93}. 
This enforces momentum conservation for the interlayer pair tunneling
   process.
(The effect of $\bk$-space broadening of the ILT kernel, \emph{e.g.}
   due to impurities, has been considered in
   Ref.~\onlinecite{Fjaerestad:98}.)

A straightforward mean-field analysis of Eq.~(\ref{eq:H}) for a
   bilayer complex and an in-plane superconducting instability in the
   $d$-wave channel yields the gap equation \cite{Angilella:99}:
\begin{equation}
\Delta_\bk = \frac{\Delta_0 g_\bk}{1-T_J (\bk)\chi_\bk} ,
\label{eq:gap1}
\end{equation}
where $\Delta_0$ is determined self-consistently from
\begin{equation}
1 = \frac{V}{N} \sum_{\bk^\prime} g_{\bk^\prime}^2
   \frac{\chi_{\bk^\prime}}{1-T_J (\bk^\prime )\chi_{\bk^\prime} }.
\label{eq:gap2}
\end{equation}
Here, $\chi_\bk = (2E_\bk )^{-1} \tanh (\beta E_\bk /2)$ is the pair
   susceptibility at the inverse temperature $\beta = (\kB T)^{-1}$,
   $E_\bk = \sqrt{\xi_\bk^2 + |\Delta_\bk |^2}$ is the upper branch of
   the superconducting spectrum, and $N$ is the number of lattice sites.

Eqs.~(\ref{eq:gap1}) and (\ref{eq:gap2}) should be immediately compared and
   contrasted with Eq.~(9) in Ref.~\onlinecite{Chen:04} (for a
   multilayered complex) and Eq.~(1) in Ref.~\onlinecite{Chen:04a}
   (for a bilayer complex).
Even without going into the subtleties of the more general derivation
   for an $n$-layered complex (for which, see
   Refs.~\onlinecite{Sudboe:94c,Angilella:99b}), or with the
   competition among several in-plane pairing channels
   (Ref.~\onlinecite{Angilella:99}), it is apparent that the gap
   function within 
   the ILT model, Eq.~(\ref{eq:gap1}), is characterized by a
   \emph{local} prefactor $[1-T_J (\bk)\chi_\bk ]^{-1}$ which, albeit
   linked self-consistently to $\Delta_0$ \emph{via}
   Eq.~(\ref{eq:gap2}), is responsible of most of the quantitative and
   qualitative features of the model.
Such a structure is missing in Refs.~\onlinecite{Chen:04,Chen:04a}.
Even though the actual symmetry of the gap function is independent of
   the ILT kernel, and is rather determined by the $d$-wave nature of
   the in-plane coupling, the ILT mechanism endows the gap function
   with a nontrivial structure in $\bk$-space \cite{Angilella:99},
   which has been shown to be consistent with ARPES results
   \cite{Angilella:00}.
Moreover, the `renormalized' pair susceptibility in the summand of
   Eq.~(\ref{eq:gap2}), \emph{viz.} $\chi_\bk \mapsto \chi_\bk /
   [1-T_J (\bk)\chi_\bk ]$, which is due to the local ILT tunneling
   amplitude, gives rise to additional, algebraic divergences
   in the energy dependence of the integrated pair susceptibility, as
   opposed to the logarithmic one, typical of BCS theory \cite{AGD}.
This is directly responsible of the enhancement of $T_c$ within the
   ILT model.
In particular, in the case of a bilayer complex, one analytically
   finds a lower bound for $T_c$ as
\begin{equation}
\kB T^\ast (\mu) =
\begin{cases}
\displaystyle
\frac{T_J}{64} \left( \frac{\mu_\perp - \mu}{\mu_\perp + 2t}
   \right)^4 , & \mu_\perp \leq \mu < \mu_{\mathrm{VH}} ,\\
\displaystyle
\frac{T_J}{64} \left( \frac{\mu_\top - \mu}{\mu_\top - 2t}
   \right)^4 , & \mu_{\mathrm{VH}} \leq \mu \leq \mu_\top ,
\end{cases}
\end{equation}
where nearest ($t$) and next-nearest ($t^\prime$) hopping have been
   assumed, and $\mu_\perp = -4t+4t^\prime$, $\mu_\top =
   4t+4t^\prime$, and $\mu_{\mathrm{VH}} = -4t^\prime$
   denote the bottom, the top of the band, and the location of the
   Van~Hove singularity, respectively \cite{Angilella:99}.

On the contrary, the ILT kernel enters Chen \emph{et al.}'s gap
   equation in Eq.~(9) of Ref.~\onlinecite{Chen:04} and Eq.~(1) of
   Ref.~\onlinecite{Chen:04} as an additional contribution to the
   non-local in-plane coupling term, \emph{i.e.} it amounts to
   defining another in-plane interaction, with no reference to
   interlayer tunneling.
\emph{This same observation applies to the general case of an
   $n$-layered complex.} 
In that case, the gap equation for each layer should also contain a
   local contribution due to the ILT mechanism between adjacent
   layers (again, absent in Ref.~\onlinecite{Chen:04}), with an ILT
   renormalized pair susceptibility $\chi_\bk^\ell / [1-T_J
   (\bk)\hat{\chi}_\bk^\ell ]$ for each layer
   \cite{Sudboe:94c,Angilella:99b}, with 
\begin{eqnarray}
\hat{\chi}_\bk^\ell &=& \left[\sin\left( \frac{\ell\pi}{n+1}
   \right)\right]^{-1} \left[ \chi_\bk^{\ell+1} \sin \left(
   \frac{(\ell+1)\pi}{n+1} \right) \right. \nonumber\\
&&+ \left. \chi_\bk^{\ell-1} \sin \left(
   \frac{(\ell-1)\pi}{n+1} \right) \right],
\end{eqnarray}
which can be further simplified in the limit of uniform hole-content
   in all layers (as is tacitly assumed in
   Ref.~\onlinecite{Chen:04}).
In analogy to the bilayer case, the condition
\begin{equation}
\min_\bk [1-T_J (\bk)\chi_\bk^\ell ] = 0
\end{equation}
then implicitly defines a lower bound $T_c^{\ast\ell}$ for the
   critical temperature 
   corresponding to the onset of superconductivity \emph{in the given
   layer $\ell$.}
Therefore, for nonuniform hole-content among inequivalent layers, as
   is the case for the multilayered cuprates
   \cite{Trokiner:91,Ambrosch-Draxl:04}, one can estimate a lower
   bound to $T_c$ as $\max_\ell T_c^{\ast\ell}$.
A nonuniform distribution of the overall hole-content among
   inequivalent layers can be conveniently described by means of
   appropriate models \cite{Angilella:99b}.
This then enables us to identify whether the superconducting instability
   first sets in in inner or outer layers.
One finds a crossover as function of the overall hole-content
   \cite{Angilella:99b}, which has been related to the observed
   `kinks' in the pressure dependence of $T_c$ in several layered
   cuprates \cite{Wijngaarden:99}.

In conclusion, we have pointed out an incorrect derivation of the gap
   equation(s) for layered cuprates within the ILT model
   \cite{Chen:04,Chen:04a} for the general case of $n$ superconducting
   layers per unit cell.
This in turn leads to a failure to capture most of the
   qualitative and quantitative features of the theory, both for
   bilayered and multilayered compounds.
As a consequence, the theoretical analysis of the high pressure data
   in Ref.~\onlinecite{Chen:04a} is not consistent with the ILT
   mechanism.
On the other hand, a more careful analysis of the ILT model
   \cite{Sudboe:94c}, when taking into account a nonuniform
   hole-content distribution among inequivalent layers
   \cite{Angilella:99b}, is indeed able to reproduce the observed
   pressure dependence of $T_c$ in multilayered cuprates
   \cite{Wijngaarden:99}.

\begin{acknowledgments}
We thank S. Chakravarty and J. O. Fj\ae{}restad for useful discussions
   and correspondence.
\end{acknowledgments}

\begin{small}
\bibliographystyle{apsrev}
\bibliography{a,b,c,d,e,f,g,h,i,j,k,l,m,n,o,p,q,r,s,t,u,v,w,x,y,z,zzproceedings,Angilella}
\end{small}

\end{document}